\documentclass[12pt,a4paper]{article}%
\usepackage{amssymb}
\usepackage{amsfonts}
\usepackage{amsmath}
\usepackage{graphicx}%
\providecommand{\U}[1]{\protect\rule{.1in}{.1in}}

\begin{document}

\title{\textbf{Stability and Identification with Optimal Macroprudential Policy
Rules\thanks{We thank for helpful comments Antoine d'Autume, Hippolyte
d'Albis, Gunther Capelle-Blancard, Robert Chirinko, Jean-Pierre Drugeon, Pedro
Garcia Duarte, Roger Farmer, Stephane Gauthier, Mauro Napoletano, Michel de
Vroey and Bertrand Wigniolle, as well as participants in the Macroeconomics in
Perspective Workshop in Louvain la Neuve and the seminar "Dynamique de la
macro\'{e}conomie" in Paris 1 Pantheon Sorbonne. } }}
\author{Jean-Bernard Chatelain\thanks{Paris School of Economics, Universit\'{e} Paris
I Pantheon Sorbonne, CES, Centre d'Economie de la Sorbonne, 106-112 Boulevard
de l'H\^{o}pital 75647 Paris Cedex 13. Email:
jean-bernard.chatelain@univ-paris1.fr} and Kirsten Ralf\thanks{ESCE
International Business School, 10 rue Sextius Michel, 75015 Paris, Email:
Kirsten.Ralf@esce.fr.}}
\maketitle

\begin{abstract}
This paper investigates the identification, the determinacy and the stability
of \textit{ad hoc}, "quasi-optimal" and optimal policy rules augmented with
financial stability indicators (such as asset prices deviations from their
fundamental values) and minimizing the volatility of the policy interest
rates, when the central bank precommits to financial stability. Firstly,
\textit{ad hoc} and quasi-optimal rules parameters of financial stability
indicators cannot be identified. For those rules, non zero policy rule
parameters of financial stability indicators are \textbf{observationally
equivalent} to rule parameters set to \textit{zero} in another rule, so that
they are \textit{unable to inform monetary policy}. Secondly, under
controllability conditions, optimal policy rules parameters of financial
stability indicators can all be identified, along with a bounded solution
\textbf{stabilizing an unstable economy} as in Woodford (2003), with
determinacy of the initial conditions of non- predetermined variables.

\textbf{JEL\ classification numbers}: C61, C62, E43, E44, E47, E52, E58.

\textbf{Keywords:} Identification, Financial Stability, Optimal Policy under
Commitment, Augmented Taylor rule, Monetary Policy.

\end{abstract}

\begin{quotation}
"\textit{If we ran the Taylor rule regression in data generated by the
new-Keynesian model, we would recover the shock autocorrelation process, not
the Taylor rule parameter}". Cochrane (2011, online appendix, p.15).

"\textit{We may omit consideration of the transversality conditions, as we
shall consider only bounded solutions to these equations, which necessarily
satisfy the transversality conditions.}" Woodford (2003, p.865).
\end{quotation}

\section{Introduction}

Should financial stability concerns influence monetary policy decisions? For a
policy-maker (Stein (2014)), the argument rests on three assumptions. First,
the Federal Reserve cares about minimizing a quadratic loss objective function
which includes a "risk" term, given by the variance of realized unemployment,
which depend on financial market vulnerability. "\textit{Second, there is some
variable summarizing financial market vulnerability which is influenced by
monetary policy... The third and final assumption is that the risks associated
with an elevated value of financial market vulnerability cannot be fully
offset at zero cost with other non-monetary tools, such as financial
regulation}" (Stein (2014), p. 2-3). In this context, the recent
macro-prudential dynamic stochastic general equilibrium (DSGE)\ literature
compares the outcomes of \textit{ad hoc} Taylor rules of the central bank "not
augmented" versus "macroprudential rules" augmented with macroprudential
indicators such as asset prices and/or households', non-financial firms' and
banks' leverage, credit spreads, liquidity ratios and so on (e.g. Beau, Clerc
and Mojon (2012), Smets (2013), Chadha, Corrado and Corrado (2013), Gambacorta
and Signoretti (2014)).

This paper investigates in a \emph{general} framework under which conditions
the augmented policy rule parameters of financial stability indicators are
\emph{identified} within \textit{ad hoc, quasi-optimal or optimal policy rules
under commitment to financial stability}\textit{. }It provides complementary
results with respect to the lack of identification of \textit{non augmented}
\textit{ad hoc} Taylor rules parameters found by Cochrane (2011), Komunjer and
Ng (2011) and Caglar, Chadha and Shibayama (2012).

With linear quadratic rational expectations optimal rules, the policy-maker
determines optimal feedback policy rule parameters as a Stackelberg leader in
a dynamic game with the private sector. She minimizes a quadratic loss
function subject to the private sector first order conditions, linearized
around an equilibrium (Woodford (2003), Blake and Kirsanova (2012), Ljungqvist
and Sargent (2012, chapter 19), Miller and Salmon (1985) among others). DSGE
models include $n$ "predetermined" variables with known initial value, such as
stationary auto-correlated shocks and capital stocks and $m$
"non-predetermined" variables which are "forward" rational expectations
variables with unknown initial values. Examples here are expected inflation,
output gap, asset prices, private credit and so on. Levine and Currie (1987)
called a policy rule "quasi-optimal" whenever it includes constraints for the
parameters of non-predetermined variables to be equal to zero. "Quasi-optimal"
policy rules are an intermediate modeling step towards time consistent policy
rules (Blake and Kirsanova (2012)).

With quasi-optimal policy rules and ad hoc rules, the policy-maker and the
private sector assume transversality conditions (there are no bubbles for
non-predetermined variables such as asset prices), seeking Blanchard and
Kahn's (1980) unique stable solution. By assumption, it follows that non-
predetermined variables are a linear function (with time invariant
coefficients) of predetermined variables for all periods of the model. Then,
if one substitutes non-predetermined variables by predetermined variables in a
policy rule with rule parameters of non predetermined variables which are
\emph{not} all equal to zero, it leads to another \emph{observationally
equivalent }policy rule with rule parameters for non predetermined variables
which are \emph{all equal to zero}. \emph{The parameters of non-predetermined
variables cannot be identified in a quasi-optimal or an ad hoc rule}. In more
technical words, the rank of the dynamic system is equal to the number $n$ of
predetermined variables and the eigenvalues related to $m$ "Jordan
transformed" non-predetermined variables have been set to zero. Hence, in
those models, discussing whether Central Bank policy makers should augment or
not Taylor rules with asset prices or financial instability indicators does
\emph{not }inform monetary policy.

If not Blanchard and Kahn's (1980) conditions, then what? This paper proposes
\emph{sufficient conditions} for the identification of rule parameters of non
predetermined variables in a general case of Woodford's (2003) and Ljungqvist
and Sargent's (2012, chapter 19) optimal policy rules under commitment. These
optimal rules are "over stable", according to Levine and Currie's (1987)
definition, as the number of stable dimensions of the dynamic system (stable
eigenvalues) is equal to $n+m$ the number of variables: it is larger than the
number $n$ of predetermined variables. The rank of the dynamic system under
control is equal to $n+m$.

Kalman (1960) defined a $\emph{controllable}$ dynamic system with linear
feedback rule when a policy-maker is able to move this dynamic system in any
state during any variation of time. A sufficient condition for a controllable
system is that the policy rule instruments can have an \emph{effect} on all
the ("Jordan" transformed) variables, in particular, all the non-predetermined
variables. This is related to Stein's (2014) second assumption: \textit{there
is some variable summarizing financial market vulnerability which is
influenced by monetary policy. }In DSGE models, a subset of a system can be
checked to be controllable. This subset excludes stationary auto-regressive
shocks which are exogenous, hence not controllable.

This paper states that if this subset of the dynamic system of a DSGE\ is
\emph{controllable} and if all its eigenvalues are \emph{distinct} and
\emph{stable}, then the linear quadratic regulator optimal policy rule
parameters of all controllable variables, including non predetermined
variables (more precisely, their predetermined shadow prices), are
\emph{unique}, this set of linear quadratic regulator rule parameters has a
one to one correspondence with the set of distinct and stable eigenvalues, and
all rule parameters can be \emph{identified} in optimal policy interest rules
under commitment \`{a} la Woodford (2003). Indeed, testing optimal rules under
commitment against quasi-optimal rules or ad hoc rules is \emph{impossible},
because the rule parameters of financial instability indicators such as asset
prices in "quasi-optimal", ad hoc and time consistent "augmented" Taylor rules
cannot be identified, and then cannot be estimable.

The paper then mentions a very important result for the determinacy of New
Keynesian models with optimal rules under commitment, \emph{which is the
opposite of the conventional determinacy criterion for ad hoc rules} (Cochrane
(2011)). The conventional determinacy criterion is the equality of the number
of stable eigenvalues to the number of predetermined variables (Blanchard and
Kahn's (1980)). The conventional view leads to the alternative: "\emph{bubbles
versus sunspots}" (Loisel (2009)). \emph{New Keynesian macroprudential DSGE ad
hoc augmented Taylor rules should not stabilize potential bubbles of }$m$
\emph{non-predetermined variables such as asset prices and private credit} in
order to maintain the \emph{uniqueness} of a knife-edge stable equilibrium
path, knowing that infinitesimal deviations from this path lead to diverging
bubbles for the $m$ non-predetermined variables. Else, there will be an
infinity of initial values ("sunspots") for the $m$ non-predetermined
variables, if ever diverging bubbles are stabilized by Old Keynesian
policy-makers rules. However, several economists consider that the knife edge
equilibrium is not unique and that an infinity of rational expectations
multiple equilibria with diverging paths (bubbles) are also valid (Burmeister
(1980), Cochrane (2011), Christiaans (2013)).

Quasi-optimal and time-consistent rules under commitment may face multiple
equilibria and indeterminacy when satisfying Blanchard and Kahn's (1980)
condition of the equality between the number of stable eigenvalues and the
number of predetermined variables (Blake and Kirsanova (2012)). If the system
is controllable with $n+m$ stable dimensions (stable eigenvalues related to
"stable" eigenvectors), there are a number of ways to find $n$ eigenvectors
related to stable eigenvalues equal to the number $n$ of predetermined
variables, satisfying Blanchard and Kahn's (1980) condition. This number is
equal to the number of subsets of $n$ distinct eigenvectors among $n+m$
eigenvectors (Blake and Kirsanova (2012)). For optimal rules under commitment,
there is only one subset of $n+m$ distinct eigenvectors of stable eigenvalues
among $n+m$ eigenvectors. Finally, if the system is controllable, the Lagrange
multipliers of non-predetermined variables with optimal rules under commitment
are equal to zero at the initial date, in order to minimize the marginal value
of the loss function. This allows to "determine" the unique initial value of
each non-predetermined variable, such as financial stability indicators.

Additionally, this paper highlights two useful properties of optimal rules
under commitment which inform monetary policy. Empirical literature documents
that unexpected changes in the nominal interest rate have a significant effect
on real stock prices and on housing prices (Challe and Giannitsarou (2014)).
However, Central Bankers fear that stabilizing asset prices and credit bubbles
may lead to too much volatility of their interest rate. A\ first issue faced
by policy-makers is then to set a trade-off between the policy rate volatility
versus targeting financial stability, understood as leaning against credit and
asset price bubbles. This issue cannot be informative with "quasi-optimal"
rules, as non-predetermined variables cannot be identified.

A\ second issue is related to the ability of macro-prudential policy to
decrease the wealth effect channel of financial instability, i.e. the
correlation between asset prices and capital. This issue cannot be addressed
with "quasi-optimal" rules, because there is a fixed exact linear relationship
(and correlation) between non-predetermined variables (asset prices) and
predetermined variables (capital) following Blanchard and Kahn (1980) assumption.

This paper proceeds as follows. Section 1 and 2\ presents identification of
rule parameters with quasi-optimal versus with optimal rules under commitment
to financial stability. Section 3\ concludes with potential extensions.

\section{Identification in "quasi-optimal" rules with pre-commitment assuming
no-bubbles}

Our analysis uses without lack of generality the deterministic setup. The
certainty equivalence property of linear quadratic optimal control models
implies that optimal rule parameters do not depend from an appropriate vector
of random shocks which can be added (Anderson \textit{et al.} (1996), Blake
and Kirsanova (2012)). The Central Bank as a Stackelberg leader
\textbf{commits} to a sequence of decision rules at time $0$, in a Ramsey
problem (Ljundqvist and Sargent (2012), chapter 19). She minimizes her loss
function by finding a sequence of decision rules $r_{t}$:%

\begin{align}
&  \underset{\left\{  r_{t},k_{t+1,}q_{t+1}\right\}  }{\max}-\frac{1}{2}%
{\displaystyle\sum\limits_{t=0}^{+\infty}}
\beta^{t}\left[
\begin{array}
[c]{c}%
\left(  \frac{\mathbf{k}_{t}-\mathbf{k}^{\ast}}{\mathbf{k}^{\ast}}\right)
^{T}\mathbf{Q}_{nn}\left(  \frac{\mathbf{k}_{t}-\mathbf{k}^{\ast}}%
{\mathbf{k}^{\ast}}\right)  +\left(  \frac{\mathbf{q}_{t}-\mathbf{q}^{\ast}%
}{\mathbf{q}^{\ast}}\right)  ^{T}\mathbf{Q}_{mm}\left(  \frac{\mathbf{q}%
_{t}-\mathbf{q}^{\ast}}{\mathbf{q}^{\ast}}\right) \\
+\left(  \frac{\mathbf{k}_{t}-\mathbf{k}^{\ast}}{\mathbf{k}^{\ast}}\right)
^{T}\mathbf{Q}_{nm}\left(  \frac{\mathbf{q}_{t}-\mathbf{q}^{\ast}}%
{\mathbf{q}^{\ast}}\right)  +\left(  \frac{\mathbf{q}_{t}-\mathbf{q}^{\ast}%
}{\mathbf{q}^{\ast}}\right)  ^{T}\mathbf{Q}_{mn}\left(  \frac{\mathbf{k}%
_{t}-\mathbf{k}^{\ast}}{\mathbf{k}^{\ast}}\right) \\
+\rho\left(  r_{t}-r^{\ast}\right)  ^{2}%
\end{array}
\right] \nonumber\\
&  =-\left(
\begin{array}
[c]{c}%
\frac{\mathbf{k}_{0}-\mathbf{k}^{\ast}}{\mathbf{k}^{\ast}}\\
\frac{\mathbf{q}_{0}-\mathbf{q}^{\ast}}{\mathbf{q}^{\ast}}%
\end{array}
\right)  ^{T}\mathbf{P}\left(
\begin{array}
[c]{c}%
\frac{\mathbf{k}_{0}-\mathbf{k}^{\ast}}{\mathbf{k}^{\ast}}\\
\frac{\mathbf{q}_{0}-\mathbf{q}^{\ast}}{\mathbf{q}^{\ast}}%
\end{array}
\right)  \text{ }%
\end{align}

The Central Bank loss function is subject to a closed loop dynamics including
the feedback rule:%

\begin{equation}
\left(
\begin{array}
[c]{c}%
\mathbf{k}_{t+1}\\
_{t}\mathbf{q}_{t+1}%
\end{array}
\right)  =\left(  \underset{\mathbf{A}}{\underbrace{\left(
\begin{array}
[c]{cc}%
\mathbf{A}_{nn} & \mathbf{A}_{nm}\\
\mathbf{A}_{mn} & \mathbf{A}_{mm}%
\end{array}
\right)  }}+\underset{\mathbf{B}}{\underbrace{\left(
\begin{array}
[c]{c}%
\mathbf{B}_{n1}\\
\mathbf{B}_{m1}%
\end{array}
\right)  }}\underset{-\mathbf{F}}{\underbrace{\left(
\begin{array}
[c]{cc}%
\mathbf{F}_{1n} & \mathbf{F}_{1m}%
\end{array}
\right)  }}\right)  \left(
\begin{array}
[c]{c}%
\mathbf{k}_{t}\\
\mathbf{q}_{t}%
\end{array}
\right)  +\mathbf{\gamma z}_{t}%
\end{equation}
where $\beta^{\prime}$ is a discount factor, $\mathbf{k}_{t}$ is an $\left(
n\times1\right)  $ vector of variables predetermined at $t$ with initial
conditions $\mathbf{k}_{0}$ given (shocks can straightforwardly be included
into this vector); $\mathbf{q}$ is an $\left(  m\times1\right)  $ vector of
variables non-predetermined at $t$; $\mathbf{z}$ is an $\left(  k\times
1\right)  $ vector of exogenous variables; $r$ is the policy interest rate,
with a linear policy feedback rule $-\mathbf{F}$ which is a $1\times\left(
n+m\right)  $ matrix;\textbf{\ }$\mathbf{Q}_{ij}$ is a $i\times j$ positive
symmetric semi-definite matrix and $\rho>0$ is a scalar (both define the
Central Bank preferences), $\mathbf{P}$ is a symmetric matrix (when
$\mathbf{Q}$ is symmetric) which provides the optimal value of the loss
function, $\mathbf{A}$ is $\left(  n+m\right)  \times\left(  n+m\right)  $
matrix, $\mathbf{B}$ is a $\left(  n+m\right)  \times1$ matrix,
$\mathbf{\gamma}$ is a $\left(  n+m\right)  \times k$ matrix, $_{t}%
\mathbf{q}_{t}$ is the agents expectations of $\mathbf{q}_{t+1}$ defined as follows:%

\begin{equation}
_{t}\mathbf{q}_{t+1}=E_{t}\left(  \mathbf{q}_{t+1}\shortmid\Omega_{t}\right)
.
\end{equation}
$\Omega_{t}$ is the information set at date $t$ (it includes past and current
values of all endogenous variables and may include future values of exogenous
variables). According to Blanchard and Kahn (1980), a \textbf{predetermined}
variable is a function only of variables known at date $t$ so that
$\mathbf{k}_{t+1}=$ $_{t}\mathbf{k}_{t+1}$ whatever the realization of the
variables in $\Omega_{t+1}$. A \textbf{non-predetermined} variable can be a
function of any variable in $\Omega_{t+1}$, so that we can conclude that
$\mathbf{q}_{t+1}=$ $_{t}\mathbf{q}_{t+1}$ only if the realization of all
variables in $\Omega_{t+1}$ are equal to their expectations conditional on
$\Omega_{t}$.

Boundary conditions for the policy-maker's first order conditions are the
given initial conditions for predetermined variables $\mathbf{k}_{0}$ and
Blanchard and Kahn (1980) hypothesis ruling out "bubbles", i.e. the
exponential growth of the expectations of $\mathbf{w=}\left(  \mathbf{k,q,z}%
\right)  $:%

\begin{equation}
\forall t\in%
\mathbb{N}
\text{,}\exists\overline{\mathbf{w}}_{t}\in%
\mathbb{R}
^{k}\text{,}\exists\theta_{t}\in%
\mathbb{R}
\text{, such that }\left\vert E_{t}\left(  \mathbf{w}_{t+1}\shortmid\Omega
_{t}\right)  \right\vert \leq\left(  1+i\right)  ^{\theta_{t}}\overline
{\mathbf{w}}_{t}\text{, }\forall i\in%
\mathbb{R}
^{+}.
\end{equation}

Following Levine and Currie (1987) definition, a policy rule which imposes
\emph{restrictions} on the parameters of the rule is "quasi-optimal" in the
sense that it is suboptimal in the general class of linear feedback rules but
optimal within its own class. More precisely, the Central Bank is looking for
such a quasi-optimal rule which imposes \emph{restrictions} on the
coefficients related to the non-predetermined variables. When the
non-predetermined variables are excluded from the policy feedback rule,
quasi-optimal rules are \emph{time consistent}, because Calvo's (1978) shadow
prices related to non predetermined variables (denoted $\mathbf{\mu
}_{\mathbf{q}}$) are no longer computed in the policy-maker's optimization.
These quasi-optimal rules lead naturally to a third type of rules:
\emph{optimal time-consistent rules without pre-commitment} related to
discretionary policy where the policy-maker recursively optimizes again at
each future period (Blake and Kirsanova (2012)).

\textbf{Theorem 1} \textbf{(Kwakernaak and Sivan (1972, p.198))}\textit{. If
the matrix pair (}$\mathbf{A}_{n+m,n+m}$\textbf{\ }$\mathbf{B}_{n+m,1}$)
\textit{is controllable, i.e. if the Kalman (1960) controllability matrix has
full rank:}
\begin{equation}
\text{rank }\left(  \mathbf{B}\text{ \ }\mathbf{AB}\text{ \ }\mathbf{A}%
^{2}\mathbf{B}\text{ \ ... \ }\mathbf{A}^{n+m-1}\mathbf{B}\right)  =n+m
\end{equation}
\ \textit{the eigenvalues of }$\mathbf{A-BF}$\textit{\ can be arbitrarily
located in the complex plane (complex eigenvalues, however, occur in complex
conjugate pairs) by choosing a policy rule matrix }$\mathbf{F}$%
\textit{\ accordingly.}

The policy-maker considers only the set of policy rules $\mathbf{F}\left(
n\right)  $ such that $\mathbf{A-BF}$ has exactly $n_{S}=n$ stable eigenvalues
equal to the number of predetermined variables, in the hope to obtain
Blanchard and Kahn (1980) unique rational expectations solution. Let
$\mathbf{M}\left(  \mathbf{F}\right)  $ be the matrix of left eigenvectors of
$\mathbf{A-BF}$ partitioned so that (indexes represent dimensions of the block matrices):%

\begin{align}
&  \left(
\begin{array}
[c]{cc}%
\mathbf{M}\left(  \mathbf{F}\right)  _{nn} & \mathbf{M}\left(  \mathbf{F}%
\right)  _{nm}\\
\mathbf{M}\left(  \mathbf{F}\right)  _{mn} & \mathbf{M}\left(  \mathbf{F}%
\right)  _{mm}%
\end{array}
\right)  \left(
\begin{array}
[c]{cc}%
\mathbf{A}_{nn}-\mathbf{B}_{n1}\mathbf{F}_{1n} & \mathbf{A}_{nm}%
-\mathbf{B}_{n1}\mathbf{F}_{1m}\\
\mathbf{A}_{mn}-\mathbf{B}_{m1}\mathbf{F}_{1n} & \mathbf{A}_{mm}%
-\mathbf{B}_{m1}\mathbf{F}_{1m}%
\end{array}
\right) \\
&  =\left(
\begin{array}
[c]{cc}%
\mathbf{\Lambda}_{nn} & \mathbf{0}_{nm}\\
\mathbf{0}_{mn} & \mathbf{\Lambda}_{mm}%
\end{array}
\right)  \left(
\begin{array}
[c]{cc}%
\mathbf{M}\left(  \mathbf{F}\right)  _{nn} & \mathbf{M}\left(  \mathbf{F}%
\right)  _{nm}\\
\mathbf{M}\left(  \mathbf{F}\right)  _{mn} & \mathbf{M}\left(  \mathbf{F}%
\right)  _{mm}%
\end{array}
\right)
\end{align}
where $\mathbf{\Lambda}_{nn}$\textbf{\ }is a $n\times n$ diagonal matrix of
stable roots, strictly lower than $1/\sqrt{\beta}$ where $\beta$ is the
discount factor of the policy maker, and $\mathbf{\Lambda}_{mm}$ is a $m\times
m$ diagonal matrix of unstable roots. Then, the unique converging path is
determined by a linear relationship between non-predetermined variables and
predetermined variables (Blanchard and Kahn (1980)):
\begin{align*}
E_{t}\mathbf{q}_{t+1}  &  =-\mathbf{N\left(  \mathbf{F}\right)  }%
_{mn}\mathbf{k}_{t+1}=-\mathbf{M}\left(  \mathbf{F}\right)  _{mm}%
^{-1}\mathbf{M}\left(  \mathbf{F}\right)  _{mn}\mathbf{k}_{t+1}\text{ and}\\
\mathbf{q}_{0}  &  =-\mathbf{N\left(  \mathbf{F}\right)  }_{mn}\mathbf{k}%
_{0}=-\mathbf{M}\left(  \mathbf{F}\right)  _{mm}^{-1}\mathbf{M}\left(
\mathbf{F}\right)  _{mn}\mathbf{k}_{0}.
\end{align*}

This describes "jumps" of non-predetermined variables to the stable manifold
generated by predetermined variables. Then, the orthogonalized
non-predetermined variables (denoted $\mathbf{q}_{t}^{^{\prime}}$) with
unstable roots are linear combinations of orthogonalized predetermined
variables with convergent eigenvalues (Blanchard and Kahn (1980) equation A6,
p.1310). The formal derivation of matrix $\mathbf{N\left(  \mathbf{F}\right)
}$ in the general case including stochastic shocks and when the generalized
Schur method is necessary is presented in McCandless (2008), section 6.8.

If the Central Bank defines a rule on both predetermined and non-predetermined
variables $\left(  \mathbf{F}_{1n},\mathbf{F}_{1m}\right)  $, this rule is
\textbf{observationally equivalent} to a rule which depends only on
predetermined variables with weights $\left(  \mathbf{F}_{1n}^{^{\prime}%
},\mathbf{0}_{1m}\right)  $ where the Central Bank imposes \emph{restrictions}
on the coefficients of the policy feedback rule, with all weights of
non-predetermined variables equal to zero. This is detailed as follows:%

\begin{align}
r_{t}-r^{\ast}  &  =-\mathbf{F}_{1n}\left(  \frac{\mathbf{k}_{t+1}%
-\mathbf{k}^{\ast}}{\mathbf{k}^{\ast}}\right)  -\mathbf{F}_{1m}\left(
\frac{\mathbf{q}_{t+1}-\mathbf{q}^{\ast}}{\mathbf{q}^{\ast}}\right) \\
&  =-\underset{\mathbf{F}_{1n}^{^{\prime}}}{\underbrace{\left(  \mathbf{F}%
_{1n}-\mathbf{F}_{1m}\mathbf{N}_{mn}\right)  }}\left(  \frac{\mathbf{k}%
_{t+1}-\mathbf{k}^{\ast}}{\mathbf{k}^{\ast}}\right)  ,\\
\mathbf{F}  &  \mathbf{=}\left(  \mathbf{F}_{1n},\mathbf{F}_{1m}\right)
=\left(  \mathbf{F}_{1n}-\mathbf{F}_{1m}\mathbf{N}_{mn},\mathbf{0}%
_{1m}\right)  .
\end{align}

The policy-maker only needs to control predetermined variables:
\[
\underset{\left\{  R_{t}\right\}  }{\max}-\frac{1}{2}%
{\displaystyle\sum\limits_{t=0}^{+\infty}}
\beta^{\prime t}\left[  \mathbf{Q}_{nn}^{^{\prime}}\left(  \frac
{\mathbf{k}_{t}-\mathbf{k}^{\ast}}{\mathbf{k}^{\ast}}\right)  ^{2}+\rho\left(
r_{t}-r^{\ast}\right)  ^{2}\right]
\]

with a reduced loss function with weights $\mathbf{Q}_{nn}^{^{\prime}}$
depending only on pre-determined variables, which are observationally
equivalent to the initial loss function depending on both non pre-determined
variables and pre-determined variables according to the following equality:%

\[
\mathbf{Q}_{nn}^{^{\prime}}=\mathbf{Q}_{nn}+\mathbf{N\left(  \mathbf{F}%
\right)  }_{nm}^{T}\mathbf{Q}_{mm}\mathbf{N\left(  \mathbf{F}\right)  }%
_{mn}+\mathbf{Q}_{nm}\mathbf{N\left(  \mathbf{F}\right)  }_{mn}%
+\mathbf{N\left(  \mathbf{F}\right)  }_{nm}^{T}\mathbf{Q}_{mn}%
\]

subject to the closed loop system of pre-determined variables:%

\begin{equation}
\mathbf{k}_{t+1}=\left(  \mathbf{A}_{nn}^{^{\prime}}-\mathbf{B}_{n1}%
\mathbf{F}_{1n}^{^{\prime}}\right)  \mathbf{k}_{t}\text{.}%
\end{equation}

According to the following equality:%
\begin{equation}
\mathbf{A}_{nn}^{^{\prime}}=\mathbf{A}_{nn}-\mathbf{A}_{nm}\mathbf{N}%
_{mn}\text{ and }\mathbf{F}_{1n}^{^{\prime}}=\mathbf{F}_{1n}-\mathbf{F}%
_{1m}\mathbf{N}_{mn}%
\end{equation}

the closed loop system of pre-determined variables is observationally
equivalent to the top half of partitioned matrices of the non-controllable
closed loop system including also non pre-determined variables:%

\begin{align}
\left(
\begin{array}
[c]{c}%
\mathbf{k}_{t+1}\\
_{t}\mathbf{q}_{t+1}%
\end{array}
\right)   &  =\left(
\begin{array}
[c]{cc}%
\mathbf{A}_{nn}-\mathbf{B}_{n1}\mathbf{F}_{1n} & \mathbf{A}_{nm}%
-\mathbf{B}_{n1}\mathbf{F}_{1m}\\
\mathbf{A}_{mn}-\mathbf{B}_{m1}\mathbf{F}_{1n} & \mathbf{A}_{mm}%
-\mathbf{B}_{m1}\mathbf{F}_{1m}%
\end{array}
\right)  \left(
\begin{array}
[c]{c}%
\mathbf{k}_{t}\\
\mathbf{q}_{t}%
\end{array}
\right)  \text{ with}\\
\left(
\begin{array}
[c]{c}%
\mathbf{k}_{t+1}\\
_{t}\mathbf{q}_{t+1}%
\end{array}
\right)   &  =\left(
\begin{array}
[c]{c}%
\mathbf{I}_{nn}\\
-\mathbf{N}_{mn}%
\end{array}
\right)  \left(
\begin{array}
[c]{c}%
\mathbf{k}_{t+1}\\
\mathbf{k}_{t+1}%
\end{array}
\right)  \text{ and }\left(
\begin{array}
[c]{c}%
\mathbf{k}_{t}\\
\mathbf{q}_{t}%
\end{array}
\right)  =\left(
\begin{array}
[c]{c}%
\mathbf{I}_{nn}\\
-\mathbf{N}_{mn}%
\end{array}
\right)  \left(
\begin{array}
[c]{c}%
\mathbf{k}_{t}\\
\mathbf{k}_{t}%
\end{array}
\right)
\end{align}

The following proposition, initially formulated for optimal policy without
pre-commitment by Blake and Kirsanova (2012), also holds for quasi-optimal
policies with commitment:

\textbf{Proposition 1: (Blake and Kirsanova (2012), p.1333). }\textit{Let the
closed loop transition matrix where the feedback policy rule }$\mathbf{F} $
\textit{depends only on predetermined variables: }$\mathbf{A-BF=}\left(
\begin{array}
[c]{cc}%
\mathbf{A}_{nn}-\mathbf{B}_{n1}\mathbf{F}_{1n} & \mathbf{A}_{nm}\\
\mathbf{A}_{mn}-\mathbf{B}_{m1}\mathbf{F}_{1n} & \mathbf{A}_{mm}%
\end{array}
\right)  $\textit{\ have all distinct eigenvalues and be diagonalizable (which
imply that the matrix pair (}$\mathbf{A}_{n+m,n+m}$\textbf{\ }$\mathbf{B}%
_{n+m,1}$) \textit{is controllable})\textit{. Let us consider the set of
policy rules }$\mathbf{F}\left(  n_{S}\right)  $\textit{\ such that
}$\mathbf{A-BF}$\textit{\ has }$n_{S}$\textit{\ stable eigenvalues (below
}$1/\sqrt{\beta}$ \textit{where} $\beta$ \textit{is the discount factor of the
policy maker)} \textit{and }$n-n_{S}$\textit{\ unstable eigenvalues. }

\textit{Case 1. For the set of policy rules }$\mathbf{F}\left(  n_{S}\right)
$\textit{\ such that }$n_{S}<n$\textit{, the number of stable eigenvalues is
strictly below the number of pre-determined variables, there is no rational
expectations equilibrium according to Blanchard and Kahn (1980).}

\textit{Case 2. For the set of policy rules }$\mathbf{F}\left(  n_{S}\right)
$\textit{\ such that }$n_{S}=n$\textit{, the number of stable eigenvalues is
strictly equal to the number of predetermined variables, there is a unique
rational expectations equilibrium according to Blanchard and Kahn (1980).}

\textit{Case 3. For the set of policy rules }$\mathbf{F}\left(  n_{S}\right)
$\textit{\ such that }$n<n_{S}\leq n+m$\textit{, there are at most }%
$\frac{n_{S}!}{n!\left(  n_{s}-n\right)  !}$\textit{\ ways of selecting }%
$n$\textit{\ stable eigenvalues and eigenvectors among a set of }$n_{s}%
$\textit{\ stable eigenvalues or eigenvectors, with parameters of the rule
related to non-predetermined variables constrained to be equal to zero
}$\mathbf{F}_{1m}=\mathbf{0}$\textit{. Hence, the matrix }$\mathbf{N}\left(
\mathbf{F}\left(  n_{S}\right)  \right)  _{mn}$\textit{\ based on these
eigenvectors is not unique. It determines the linear relationship between non
pre-determined variables to pre-determined variables: }%

\[
E_{t}\mathbf{q}_{t+1}=-\mathbf{N\left(  \mathbf{F}\right)  }_{mn}%
\mathbf{k}_{t+1}=-\mathbf{M}\left(  \mathbf{F}\right)  _{mm}^{-1}%
\mathbf{M}\left(  \mathbf{F}\right)  _{mn}\mathbf{k}_{t+1}%
\]

\textit{which are solutions of a particular non-symmetric Riccati matrix
equation.}

As emphasized by Blake and Kirsanova (2012), multiple equilibria may remain
unnoticed. Choosing particular initial conditions for a DSGE\ may lead to
convergence to a particular equilibrium using the currently available software
programming Blanchard and Kahn (1980) solutions, without revealing the
indeterminacy of the matrix $\mathbf{N\left(  \mathbf{F}\right)  }_{mn}%
$\textit{.}

\textbf{Proposition 2. }\textit{The following results hold for quasi-optimal
rules if Blanchard and Kahn \ [1980] condition holds, that is, the number of
non pre-determined variables is equal to the number of unstable eigenvalues of
the controlled system with the transition matrix }$\mathbf{A-BF}=\left(
\begin{array}
[c]{cc}%
\mathbf{A}_{nn}-\mathbf{B}_{n1}\mathbf{F}_{1n} & \mathbf{A}_{nm}\\
\mathbf{A}_{mn}-\mathbf{B}_{m1}\mathbf{F}_{1n} & \mathbf{A}_{mm}%
\end{array}
\right)  .$

\textit{For a given matrix} $\mathbf{N}_{mn}$ \textbf{and} \ \textit{if the
matrix pair (}$\mathbf{A}_{nn}^{\prime}$\textbf{\ }$\mathbf{B}_{n1}^{\prime}%
$\textit{) is controllable, i.e. if the following Kalman (1960)
controllability matrix has full rank:}
\begin{equation}
\text{rank }\left(  \mathbf{B}_{nn}^{\prime}\text{ \ }\mathbf{A}_{nn}^{\prime
}\mathbf{B}_{1n}^{\prime}\text{ \ }\mathbf{A}_{nn}^{\prime2}\mathbf{B}%
_{n1}^{\prime}\text{ \ ... \ }\mathbf{A}_{nn}^{\prime n-1}\mathbf{B}%
_{n1}^{\prime}\right)  =n
\end{equation}
\textit{\ } \textit{the following results hold:}

\textit{1. There is a unique symmetric positive semi-definite solution
}$\mathbf{P}$\textit{\ to the discrete algebraic Ricatti equation:}%
\begin{equation}
\mathbf{P}^{\prime}=\mathbf{Q}^{\prime}+\beta\mathbf{A}^{\prime T}%
\mathbf{P}^{\prime}\mathbf{A}^{\prime}-\beta\mathbf{A}^{\prime T}%
\mathbf{P}^{\prime}\mathbf{B}^{\prime}\left(  \rho+\beta\mathbf{B}^{\prime
T}\mathbf{P}^{\prime}\mathbf{B}^{\prime}\right)  ^{-1}\beta\mathbf{B}^{\prime
T}\mathbf{P}^{\prime}\mathbf{A}^{\prime}.
\end{equation}

\textit{2. The restricted policy rule parameters of pre-determined variables
}$\mathbf{F=}\left(  \mathbf{F}_{1n}^{^{\prime}},\mathbf{0}_{1m}\right)
$\textit{\ are uniquely determined from equation:}%

\begin{equation}
\mathbf{F}_{1n}^{^{\prime}}=\beta\left(  \rho+\mathbf{B}^{\prime T}%
\mathbf{P}^{\prime}\mathbf{B}^{\prime}\right)  ^{-1}\mathbf{B}^{\prime
T}\mathbf{P}^{\prime}\mathbf{A}^{\prime}.
\end{equation}

\textit{3. Non identification of quasi-optimal rule parameters of non
pre-determined variables (such as macro-prudential risk variables) if ever
those rule parameters are distinct from zero:}%

\[
\mathbf{F}\mathbf{=}\left(  \mathbf{F}_{1n},\mathbf{F}_{1m}\right)  =\left(
\mathbf{F}_{1n}-\mathbf{F}_{1m}\mathbf{N}_{mn},\mathbf{0}_{1m}\right)  .
\]

\textit{4. Indeterminacy. According to proposition 1, there may be at most
}$\frac{\left(  n+m\right)  !}{n!m!}$\textit{\ multiple equilibria providing a
matrix }$\mathbf{N\left(  \mathbf{F}\right)  }_{mn}$:{}%

\[
E_{t}\mathbf{q}_{t+1}=-\mathbf{N\left(  \mathbf{F}\right)  }_{mn}%
\mathbf{k}_{t+1}\text{ and }E_{t}\mathbf{q}_{0}=-\mathbf{N\left(
\mathbf{F}\right)  }_{mn}\mathbf{k}_{0}%
\]

\textit{5. Bounded solution.} \textit{For the optimal policy rule }%
$\mathbf{F}_{1n}^{^{\prime}}$\textit{\ all eigenvalues of the closed loop
transition matrix }$\mathbf{A}_{nn}^{^{\prime}}-\mathbf{B}_{n1}\mathbf{F}%
_{1n}^{^{\prime}}$\textit{\ are strictly less than }$1/\sqrt{\beta}$
\textit{in absolute value}:\textit{\ }$\left\vert \lambda_{i,\mathbf{A}%
_{nn}^{\prime}\mathbf{-B}_{n1}\mathbf{F}_{1n}^{\prime}}\right\vert
<1/\sqrt{\beta}$, $1\leq i\leq n$\textit{\ . It follows that }$\underset
{t\rightarrow+\infty}{\lim}\frac{\mathbf{k}_{t}}{\left(  \sqrt{\beta}\right)
^{t}}$\textit{. Then, non-predetermined variables are bounded because }%
$_{t}\mathbf{q}_{t+1}=-\mathbf{N}_{mn}\mathbf{k}_{t+1}$.

\textit{6. Minimal volatility of the policy interest rate (}$\rho
>0,\mathbf{Q}=\mathbf{0}$\textit{). In this case, the policy rule parameters
are all equal to zero. All the (stable) eigenvalues of the open loop system
}$\mathbf{A}_{nn}^{^{\prime}}$\textit{\ remain unchanged in the closed loop
system }$\left\vert \lambda_{i,\mathbf{A-BF}}\right\vert =\left\vert
\lambda_{i,\mathbf{A}}\right\vert <1$ \textit{(Rojas (2011))}.

\textit{7. Inability of the policies to change the covariances matrix between
pre-determined and non pre-determined variables when }$\mathbf{Q}_{mn}%
\neq\mathbf{0}$\textit{\ and }$\mathbf{Q}_{nm}\neq\mathbf{0}$\textbf{,
}\textit{as it is fixed according to Blanchard and Khan (1980) condition:
}$E_{t}\mathbf{q}_{t+1}=-\mathbf{N\left(  \mathbf{F}\right)  }_{mn}%
\mathbf{k}_{t+1}$\textit{.}

\textit{8. Time consistency \`{a} la Calvo (1978). As the quasi-optimal policy
rule excludes non-predetermined variables from the optimal control problem,
the Lagrange multipliers related to there variables (at the origin of time
inconsistency issues) do not show up in the optimization.}\textbf{\ }

Most of recent macro-prudential DSGE models assume simultaneously (1) ad hoc
Taylor rules augmented by non pre-determined variables such as asset prices
and private credit and compares them with non augmented Taylor rules, (2) no
asset price bubbles and no Ponzi game condition for credit and asset prices
with Blanchard and Kahn (1980) condition. These macro-prudential DSGE models
face the same identification problem as \textit{"quasi-optimal" rules with
pre-commitment. }When analyzing the identification of all DSGE\ parameters and
not only the rule parameters, the controllability hypothesis is also
instrumental in several demonstrations in Komunjer and Ng (2011) appendix.
Rule parameters are usually found to be in the list of least identified
parameters for specific DSGE identification analysis, while auto-regressive
components of shocks are in the list of best identified parameters, see e.g.
Caglar, Chadha and Shibayama (2012).

\section{Identification and "over stable" rules with Central Bank
pre-commitment to financial stability}

Proposition 3\ summarizes the \emph{identification, determinacy and stability}
properties of "over stable" rational expectations optimal rules (Levine and
Currie (1987) terminology) with Central Bank pre-commitment to financial
stability, where the number of stable eigenvalues is larger than the number of
pre-determined variables. Woodford's (2003) famous paper on Central Bank
interest smoothing rules are "over stable" rational expectations optimal
policy interest rate rules under commitment.

Ljungqvist and Sargent (2012, chapter 19) describe a four step algorithm for
solving the optimal policy under commitment. "\textit{Step 1\ seems to
disregard the forward looking aspect of the problem. If we temporarily ignore
the fact that the} $\mathbf{q}_{0}$ \textit{component of the state}
$\mathbf{y}_{0}=\left(
\begin{array}
[c]{c}%
\frac{\mathbf{k}_{0}-\mathbf{k}^{\ast}}{\mathbf{k}^{\ast}}\\
\frac{\mathbf{q}_{0}-\mathbf{q}^{\ast}}{\mathbf{q}^{\ast}}%
\end{array}
\right)  $ \textit{is not actually a state vector, then superficially the
Stackelberg problem has the form of an optimal linear regulator problem"
}(Ljungqvist and Sargent (2012, chapter 19, p.769). Step one obtains the
matrix $\mathbf{P}$ giving the optimal value of the loss function as a
solution a matrix Riccati equation and the optimal parameters of the feedback
rule $\mathbf{F}$ "\textit{as if}" $\mathbf{q}_{t}$ are pre-determined
variables. Step 2\ seeks an "over stable" solution, that is a stabilizing
solution for the $\mathbf{y}_{t}$ \textit{including} all non-predetermined variables:%

\begin{equation}
\sum_{t=0}^{+\infty}\beta^{t}\mathbf{y}_{t}^{T}\mathbf{y}_{t}<+\infty\text{,}%
\end{equation}
solving the Lagrangian:%

\begin{equation}
-\frac{1}{2}%
{\displaystyle\sum\limits_{t=0}^{+\infty}}
\beta^{t}\left[
\begin{array}
[c]{c}%
\left(  \frac{\mathbf{k}_{t}-\mathbf{k}^{\ast}}{\mathbf{k}^{\ast}}\right)
^{T}\mathbf{Q}_{nn}\left(  \frac{\mathbf{k}_{t}-\mathbf{k}^{\ast}}%
{\mathbf{k}^{\ast}}\right)  +\left(  \frac{\mathbf{q}_{t}-\mathbf{q}^{\ast}%
}{\mathbf{q}^{\ast}}\right)  ^{T}\mathbf{Q}_{mm}\left(  \frac{\mathbf{q}%
_{t}-\mathbf{q}^{\ast}}{\mathbf{q}^{\ast}}\right) \\
+\left(  \frac{\mathbf{k}_{t}-\mathbf{k}^{\ast}}{\mathbf{k}^{\ast}}\right)
^{T}\mathbf{Q}_{nm}\left(  \frac{\mathbf{q}_{t}-\mathbf{q}^{\ast}}%
{\mathbf{q}^{\ast}}\right)  +\left(  \frac{\mathbf{q}_{t}-\mathbf{q}^{\ast}%
}{\mathbf{q}^{\ast}}\right)  ^{T}\mathbf{Q}_{mn}\left(  \frac{\mathbf{k}%
_{t}-\mathbf{k}^{\ast}}{\mathbf{k}^{\ast}}\right) \\
+\rho\left(  r_{t}-r^{\ast}\right)  ^{2}+2\beta\mathbf{\mu}_{t+1}^{T}\left(
\mathbf{Ay}_{t}+\mathbf{B}\left(  r_{t}-r^{\ast}\right)  -\mathbf{y}%
_{t+1}\right)
\end{array}
\right]  \text{ }%
\end{equation}
where $2\beta^{^{\prime}}\mathbf{\mu}_{t+1}$ is the Lagrange multiplier
associated with the linear constraint. First order conditions with respect to
$r_{t}$ and $\mathbf{y}_{t}$, respectively, are:%

\begin{align}
0  &  =\rho\left(  r_{t}-r^{\ast}\right)  +\beta\mathbf{B}^{T}\mathbf{\mu
}_{t+1}\\
\mathbf{\mu}_{t}  &  =\mathbf{Qy}_{t}+\beta\mathbf{A}^{T}\mathbf{\mu}_{t+1}%
\end{align}

Step 3 uses the property that a stabilizing solution satisfies:
\[
\mathbf{\mu}_{t}=\left(
\begin{array}
[c]{c}%
\mathbf{\mu}_{\mathbf{k},t}\\
\mathbf{\mu}_{\mathbf{q},t}%
\end{array}
\right)  =\mathbf{P}\left(
\begin{array}
[c]{c}%
\mathbf{k}_{t}\\
\mathbf{q}_{t}%
\end{array}
\right)  =\left(
\begin{array}
[c]{cc}%
\mathbf{P}_{nn} & \mathbf{P}_{nm}\\
\mathbf{P}_{mn} & \mathbf{P}_{mm}%
\end{array}
\right)  \left(
\begin{array}
[c]{c}%
\mathbf{k}_{t}\\
\mathbf{q}_{t}%
\end{array}
\right)  \text{,}\forall t\in%
\mathbb{N}%
\]

Then, predetermined variables, the optimal policy rule $\mathbf{\Phi}$ and the
closed loop system can be written as a function of predetermined variables
$\left(  \mathbf{k}_{t},\mathbf{\mu}_{\mathbf{q},t}\right)  $:%

\begin{align*}
\mathbf{q}_{t}  &  =\left(
\begin{array}
[c]{cc}%
-\mathbf{P}_{mm}^{-1}\mathbf{P}_{mn} & \mathbf{P}_{mm}^{-1}%
\end{array}
\right)  \left(
\begin{array}
[c]{c}%
\mathbf{k}_{t}\\
\mathbf{\mu}_{\mathbf{q},t}%
\end{array}
\right)  \text{ and }\mathbf{q}_{0}=-\mathbf{P}_{mm}^{-1}\mathbf{P}%
_{mn}\mathbf{k}_{0}\text{ if }\mathbf{\mu}_{\mathbf{q},t=0}=0\\
r_{t}  &  =\mathbf{\Phi}\left(
\begin{array}
[c]{c}%
\mathbf{k}_{t}\\
\mathbf{\mu}_{\mathbf{q},t}%
\end{array}
\right)  =-\mathbf{F}\left(
\begin{array}
[c]{cc}%
\mathbf{I}_{nn} & \mathbf{0}_{nm}\\
-\mathbf{P}_{mm}^{-1}\mathbf{P}_{mn} & \mathbf{P}_{mm}^{-1}%
\end{array}
\right)  \left(
\begin{array}
[c]{c}%
\mathbf{k}_{t}\\
\mathbf{\mu}_{\mathbf{q},t}%
\end{array}
\right) \\
\left(
\begin{array}
[c]{c}%
\mathbf{k}_{t+1}\\
\mathbf{\mu}_{\mathbf{q},t+1}%
\end{array}
\right)   &  =\left(
\begin{array}
[c]{cc}%
\mathbf{I}_{nn} & \mathbf{0}_{nm}\\
\mathbf{P}_{mn} & \mathbf{P}_{mm}%
\end{array}
\right)  \left(  \mathbf{A}-\mathbf{BF}\right)  \left(
\begin{array}
[c]{cc}%
\mathbf{I}_{nn} & \mathbf{0}_{nm}\\
-\mathbf{P}_{mm}^{-1}\mathbf{P}_{mn} & \mathbf{P}_{mm}^{-1}%
\end{array}
\right)  \left(
\begin{array}
[c]{c}%
\mathbf{k}_{t}\\
\mathbf{\mu}_{\mathbf{q},t}%
\end{array}
\right)
\end{align*}

To interpret empirical evidence about interest rate smoothing in the United
States, Woodford (2003) eliminates the implementation Lagrange multipliers
$\mathbf{\mu}_{\mathbf{q},t}$ in order to express the policy rule
$\mathbf{\Phi}$ as a history-dependent representation of the policy
rule\textit{\ }denoted\textit{\ }$\mathbf{\Psi}$ depending on the
variables\textit{\ }$\left(  r_{t-1,}\mathbf{k}_{t},\mathbf{k}_{t-1}\right)  $.

\textbf{Proposition 3. }\textit{If the matrix pair (}$\mathbf{A}_{n+m,n+m}%
$\textbf{\ }$\mathbf{B}_{n+m,1}$\textit{) is controllable, i.e. if the Kalman
(1960) controllability matrix has full rank:}
\begin{equation}
\text{rank }\left(  \mathbf{B}\text{ \ }\mathbf{AB}\text{ \ }\mathbf{A}%
^{2}\mathbf{B}\text{ \ ... \ }\mathbf{A}^{n+m-1}\mathbf{B}\right)  =n+m
\end{equation}
\textit{\ } \textit{the following results hold:}

\textit{1. There is a unique symmetric positive semi-definite solution
}$\mathbf{P}$\textit{\ to the discrete algebraic Ricatti equation:}%
\begin{equation}
\mathbf{P}=\mathbf{Q}+\beta\mathbf{A}^{T}\mathbf{PA}-\beta\mathbf{A}%
^{T}\mathbf{PB}\left(  \rho+\beta\mathbf{B}^{T}\mathbf{PB}\right)  ^{-1}%
\beta^{^{\prime}}\mathbf{B}^{T}\mathbf{PA}.
\end{equation}

\textit{2. Uniqueness of the policy rule parameters of all controllable
(pre-determined and non pre-determined) variables }$\mathbf{F}$\textit{\ which
is determined from equation:}%

\begin{equation}
\mathbf{F}=\beta\left(  \rho+\mathbf{B}^{T}\mathbf{PB}\right)  ^{-1}%
\mathbf{B}^{T}\mathbf{PA}.
\end{equation}
\textit{\ }

\textit{3. Identification of optimal rule parameters. If the closed loop
transition matrix }$\mathbf{A-BF}$ \textit{has all distinct eigenvalues
denoted }$\lambda_{i,\mathbf{A-BF}}$ for $1\leq i\leq n+m$, \textit{there is a
unique pole placement relationship between the set of distinct eigenvalues of
the closed loop matrix }$\lambda_{i,\mathbf{A-BF}}$ \textit{and the unique
solution of the set of parameters of the "as if }$\mathbf{q}$ \textit{is
predetermined" linear quadratic regulator policy rule }$\mathbf{F} $ \textit{.
All coefficients of the "as if }$\mathbf{q}$ \textit{is predetermined"}
\textit{policy rule }$\mathbf{F}$\textit{\ - related to both predetermined
variables and non-predetermined variables - can be identified. Then, if
}$\mathbf{P}_{mm}$ \textit{is invertible, the parameters of rule
}$\mathbf{\Phi}$ \textit{applied on the pre-determined variables }$\left(
\mathbf{k}_{t},\mathbf{\mu}_{\mathbf{q},t}\right)  $\textit{\ are also all
identified. By contrast, the parameters of the history-dependent
representation of the policy rule }$\mathbf{\Psi}$ \textit{applied on the
variables }$\left(  r_{t-1,}\mathbf{k}_{t},\mathbf{k}_{t-1}\right)  $
\textit{are usually not identified as their total number }$2n+1$\textit{\ may
differ from the number of distinct eigenvalues equal to the number of state
variables }$n+m$.

\textit{4. Determinacy. Kalman's controllability condition is a precondition
for assuming that the Lagrange multipliers related to non predetermined
variables should be all equal to zero at the initial date }$\mu_{\mathbf{q,}%
t=0}=0$\textit{\ (Bryson and Ho (1975), p.55-59; Xie (1997) provides a counter
example where Kalman's controllability condition is not satisfied). As the
Lagrange multipliers are related to the optimal value function matrix as
follows: }$\mathbf{\mu}_{\mathbf{z,}t}=\mathbf{Pz}_{t},$\textit{the initial
values of non-predetermined variables are linear functions of the initial
values of predetermined variables (Ljundqvist and Sargent's (2012, Chapter
19), Jensen (2011)):}%

\begin{equation}
\mathbf{q}_{0}=-\mathbf{P}_{mm}^{-1}\mathbf{P}_{mn}\mathbf{k}_{0}\text{ if
}\mathbf{\mu}_{\mathbf{q},t=0}=0.\text{ }%
\end{equation}

\textit{5. Bounded solution.} \textit{For this policy rule }$\mathbf{F}%
$\textit{\ all eigenvalues of the closed loop transition matrix }%
$\mathbf{A-BF}$\textit{\ (defining the evolution of the system under control)
are strictly less than }$1/\sqrt{\beta}$\textit{\ in absolute value. It
follows that }$\underset{t\rightarrow+\infty}{\lim}\frac{\mathbf{k}_{t}%
}{\left(  \sqrt{\beta}\right)  ^{t}}=\underset{t\rightarrow+\infty}{\lim}%
\frac{E_{t-1}\mathbf{q}_{t}}{\left(  \sqrt{\beta}\right)  ^{t}}=0$\textit{.
Thus the policy reaction function ensures a finite loss and we may omit
consideration of the Blanchard and Kahn's (1980) conditions on no bubbles on
non-predetermined and predetermined variables (Levine and Currie's (1987)
"over-stable" feedback rule and Woodford (2003)).}

\textit{6. Minimal volatility of the policy interest rate (}$\rho
>0,\mathbf{Q}=\mathbf{0}$\textit{). It is such that stable eigenvalues of the
open loop system are the same as in the closed loop system }$\left\vert
\lambda_{i,\mathbf{A-BF}}\right\vert =\left\vert \lambda_{i,\mathbf{A}%
}\right\vert <1$ \textit{and that unstable eigenvalues (indexed by }$i\prime
$\textit{) of the open loop system are mirrored by stable eigenvalues in the
closed loop system having their modulus such that} $\left\vert \lambda
_{i^{\prime},\mathbf{A-BF}}\right\vert =1/\left\vert \lambda_{i^{\prime
},\mathbf{A}}\right\vert <1$ \textit{(Rojas (2011))}.

\textit{7. Ability of policies to decrease the covariances matrix between
pre-determined and non pre-determined variables when the policy maker
preferences are such that }$\mathbf{Q}_{mn}\neq\mathbf{0}$\textit{\ and
}$\mathbf{Q}_{nm}\neq\mathbf{0}$\textbf{.}

\textit{8. Time inconsistency \`{a} la Calvo (1978). When the system is
controllable and without a pre-commitment constraint, a policy maker who
optimize again on period }$t+1$\textit{ would choose an initial condition
}$\mathbf{\mu}_{\mathbf{q},t+1}=\mathbf{0}$ \textit{instead of the optimal
path }$\mathbf{\mu}_{\mathbf{q},t+1}\neq\mathbf{0}$ \textit{decided on date}
$t$\textit{.}\textbf{\ }\textit{The system remains bounded and stable if ever
the policy maker chooses} $\mathbf{\mu}_{\mathbf{q},t+k}=\mathbf{0}$\textbf{
}\textit{on all following periods (}$k>1$\textit{): the policy maker is not
time inconsistent in the sense of promoting the instability of pre-determined
variables, instead of their stability.}\textbf{\ }

\section{Conclusion}

Parameters of financial instability variables in optimal policy rules with
Central Bank commitment to financial stability can be identified, whereas it
is not the case for "quasi-optimal" rules. Moreover, Kalman's (1960)
controllability condition is a sufficient condition for determinacy and
stability of these optimal policy rule under commitment.

Many extensions of optimal rules under commitment to financial stability are
feasible. Firstly, \emph{all} existing macro-prudential DSGE papers including
an \textit{ad hoc} augmented Taylor rules can be extended with additional
simulations of optimal rules under commitment. These simulations and
estimations would then be compared with the ones of \textit{ad hoc} augmented
Taylor rules. Secondly, \emph{robust optimal control} \`{a} la Hansen and
Sargent (2008), where the Central Bank optimal policy intends to minimize the
worst of outcomes where they do not know with certainty fundamentals (as in
Lorenzoni (2010)), is close to nowadays policy makers concerns.

Finally, the parameters of the Central Bank loss function should be
endogenously determined. They result from a delegation problem taking into
account the bargaining powers of the private sector divergent interests
between lenders versus borrowers and between the banking sector versus
non-financial sectors. Big banks benefit from financial instability over the
business cycle through higher returns and larger informational rents during
booms while being bailed out in case of distress. More precisely, the legal
institutions surrounding the pre-commitment to a financial stability mandate
should be investigated. Not only the Central Bank should be independent from
government, but also it should be \emph{independent} from the private banking sector.

\end{document}